\documentclass[journal, letterpaper]{IEEEtran}
\pdfoutput=1

\usepackage[cmex10]{amsmath}
\usepackage{amsfonts}
\usepackage[pdftex]{graphicx}
\usepackage{import}
\usepackage{color}
\usepackage{hyperref}
\usepackage{flushend}
\newcommand{\cotwelve}{$^\text{12}$CO(1-0) }
\newcommand{\cothirteen}{$^\text{13}$CO(1-0) }

\begin{document}

\title{The CAMbridge Emission Line Surveyor (CAMELS)}

\author{Christopher~N.~Thomas,~Stafford~Withington,~Roberto~Maiolino,~David~J.~Goldie, Eloy~de Lera Acedo,~Jeff~Wagg,~Ray~Blundell,~Scott~Paine~and~Lingzhen~Zeng 
\thanks{C.N. Thomas, S. Withington, R. Maiolino, D. Goldie, E. de Lera Acedo and J. Wagg are with the Department of Physics, Cambridge University, Cambridge, CB3 0HE, UK.  R. Blundell, S. Paine and L. Zeng are with the Harvard-Smithsonian Center for Astrophysics, 60 Garden Street, Cambridge, MA 02138, USA.  Corresponding author C.N. Thomas (c.thomas@mrao.cam.ac.uk)} }

\maketitle

% Uncomment for draft
%\onehalfspacing

\begin{abstract}
The CAMbridge Emission Line Surveyor (CAMELS) is a pathfinder program to demonstrate on-chip spectrometry at millimetre wavelengths. CAMELS will observe at frequencies from 103--114.7\,GHz, providing 512 channels with a spectral resolution of R = 3000.  In this paper we describe the science goals of CAMELS, the current system level design for the instrument and the work we are doing on the detailed designs of the individual components.  In addition, we will discuss our efforts to understand the impact that the design and calibration of the filter bank on astronomical performance.  The shape of the filter channels, the degree of overlap and the nature of the noise all effect how well the parameters of a spectral line can be recovered.  We have developed a new and rigorous method for analysing performance, based on the concept of Fisher information.  This can in be turn coupled to a detailed model of the science case, allowing design trade-offs to be properly investigated.
\end{abstract}

\section{Introduction}

The aim of the CAMbridge Emission Line Surveyor (CAMELS) project is to provide an operational demonstration of an `on-chip' spectrometer for W-band astronomical observations.  The basic principle of on-chip spectrometry is to provide the spectral channelisation on the same wafer as the detectors, rather than via additional optics, as in the case of a Fourier Transform Spectrometer (FTS) or a grating spectrometer. A promising approach is the Integrated Filter-Bank Spectrometer (IFBS), where a set of narrow-band electrical filters are used to disperse the signal into spectral channels before detection.  The antenna, filter-bank and detectors can all be fabricated together on the same chip, with the possibility of integrating many such systems on a single wafer to realise a large spectroscopic imaging array.  The resulting solution is significantly more compact than a grating or FTS design, and is mechanically more rugged.  Although it is unlikely an IFBS can provide the same spectral resolution as a coherent receiver, it can provide a much larger instantaneous fractional bandwidth, which is limited only by the number of detectors that can be read out simultaneously.  Additionally, an IFBS can offer greater sensitivity than a coherent receiver via the use of superconducting direct detectors, which avoid the quantum noise limits of mixers. 

Large format imaging arrays or multi-object systems that use on-chip spectrometry to provide moderate spectral resolution (R $<$ 3000) over a large bandwidth are viewed as a transformative technology for astronomy at wavelengths from the millimetre right through to the far infrared.  A key envisaged use is as a survey instrument, providing CO detections over extended redshift ranges on a scale comparable to the Sloan Digital Sky Survey.  Such a dataset would yield unbiased redshift surveys, as well as an unprecedented amount of information about the evolution of the molecular gas content in galaxies throughout the cosmic epochs.  A large format spectroscopic imaging array would be useful for intensity mapping experiments on CO and [CII]\,\cite{gong2012intensity}, as well as for quickly obtaining maps of the spatial distribution of different gas species such as HCN, SiO and CN in local galaxies.  In addition, the combination of high sensitivity and broad bandwidth is particularly suited to the detection of the broad wings of the molecular transitions, which will make it possible to survey massive molecular outflows in galaxies\,\cite{cicone2012physics}.  

There are two parts to the CAMELS project.  The first is the development of the necessary technologies to build an IFBS at these frequencies, including filter and detector designs.  The second is the deployment of a pathfinder instrument on a telescope, so as to explore the operational issues associated with an IFBS.  These include flux- and frequency- calibration, operation in varying backgrounds and the corresponding development of optimal observing strategies.  The operational demonstration is being made possible by a collaboration between Cambridge University and the Harvard Smithsonian Center for Astrophysics.  It is planned that the prototype will be tested on the Greenland Telescope (GLT) during its 18 month commissioning phase at Thule Air Force Base, beginning in 2016.  

The pathfinder instrument will target \cotwelve($\nu_0$ = 115.271\,GHz) and \cothirteen ($\nu_0$ = 110.201\,GHz) line emission from galaxies, using two pairs of spectrometer pixels observing in the frequency range from 103\,GHz to 114.7\,GHz.  One pair will observe in the range 103--109.8\,GHz and the other 109.8--114.7\,GHz, with each pixel providing 256 spectral channels with spectral resolution $R = \Delta \nu / \nu$ = 3000 (a velocity resolution of 100\,km/s).  The two different sub-bands test very different observing regimes. The lower sub-band is well away from the edge of the atmospheric window, so background loading is low.  However, the sub-band contains the faint emission from the more distant galaxies.  The upper sub-band, being nearer the edge of the atmospheric window, suffers higher background loading from the atmospheric oxygen absorption line at 119\,GHz.  However, it the brighter emission from local galaxies that will be observed.  The two sub-bands will therefore allow the study of the technology's performance for both detecting faint lines in low background, and also mapping bright line emission in a strong, variable, background.  The noise behaviour in the two regimes is, for example, expected to be different, and this may influence the design of the filter bank (see Section \ref{sec:spectrometer_model}).  As another example, the KIDs for the upper sideband will need to be able to cope with higher background, and the effect of dynamic variation of this loading during an observation need to be understood.  Having two spatial pixels observing within each sub-band will allow an investigation of systematic effects via the comparison of two notionally identical units, as well as chopping strategies.  A number of other on-chip-spectrometers demonstration programs are also underway, including DESHIMA\cite{endo2012development}, SuperSpec\cite{shirokoff2012mkid} and MicroSpec.  Together, these instruments cover the observing range 300\,GHz--1\,THz, so CAMELS provides a highly complementary demonstration of the technology at lower frequencies. 

Although the primary purpose of CAMELS program is as a technology pathfinder, it has also been designed so as to provide significant scientific return.  The planned observing program will allow the measurement of the line emission from a significant sample of galaxies in the redshift ranges 0.05--0.13 ($^\text{12}$CO) and 0.003--0.961 ($^\text{13}$CO).  We estimates CAMELS should be able to survey approximately 1000 galaxies over the planned demonstration phase, which would be an order of magnitude improvement over existing surveys.  This data will be used to investigate the dependence of the Schmidt-Kennicutt star-formation on gas environment and metallicity.  In particular, around 10\% of the sample is predicted to be detected in both \cotwelve and \cothirteen\!.  These double detections will enable further constraints to be put on the CO--H$_\text{2}$ conversion factor in different galaxies classes.  Additionally, for the closest galaxies (z $\approx$ 0.005), the size of the GLT beam ($\approx$\,50'' at 100\,GHz) is sufficiently small to allow mapping of the gas distribution in the galaxy on the scale of $\approx$\,5\,kpc.  

This paper will give a brief overview of the design of CAMELS instrument as currently envisaged.  In Section \ref{sec:instrument_overview} we outline the system level design of the instrument and the intended science goals, addressing the reasoning behind some of our major choices.  Section \ref{sec:optical_coupling} will describe our work on the optical coupling 70--110\,GHz radiation into a KID.  This is a novel aspect of the design as, at the time of writing, a KID that is sensitive at these wavelengths has not been demonstrated experimentally.  Finally, in Section \ref{sec:filter_bank_performance} we will discuss some theoretical work we have undertaken that looks at finding a performance metric by which to compare different filter bank designs.

\section{Instrument Overview}\label{sec:instrument_overview}

\subsection{System Level Design}

\begin{figure*}
\centering
\includegraphics[width=16cm]{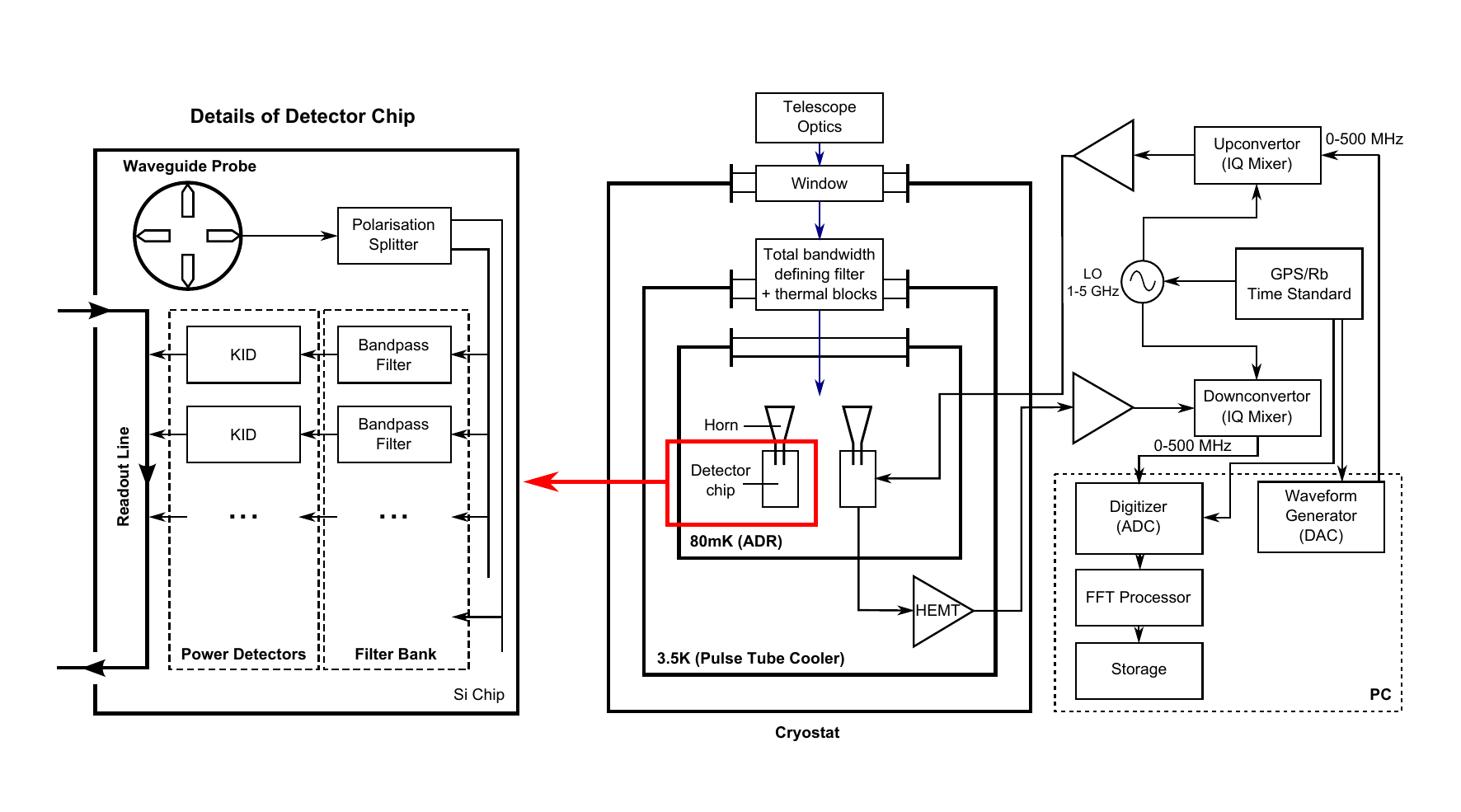}
\caption{System diagram for CAMELS.  The left-hand side of the image shows the layout of the detector chips, while the right-hand side shows details of the cryostat and the readout system.}
\label{fig:camels_system_diagram} 
\end{figure*}

Figure \ref{fig:camels_system_diagram} shows the current system level design for CAMELS.  The basic architecture is similar to that of other IFBS projects, such as DESHIMA\cite{endo2012development} and SuperSpec\cite{shirokoff2012mkid}.  The incident radiation is coupled onto a transmission line on the chip by an antenna, where it is fed into a filter bank circuit.  The filter bank splits the signal into narrow frequency bands, each of which appears at a different output and is sensed by a separate power detector.

CAMELS will use Kinetic Inductance Detectors (KIDs) for the power detection stage, principally for the ease with which a large number can be frequency multiplexed together onto a single readout line.  At present we plan to use a relativity conventional KID design, based on a quarter-wave resonator readout at gigahertz frequencies.  In the KID community there is currently a movement towards lumped-element designs that can be read out at much lower frequencies (on the order of 100\,MHz), which helps reduce two-level system noise and improves the multiplexing factor.  We have chosen a well-studied design so as to reduce risk, and allow us to focus on the optical coupling scheme and demonstrating the filter bank technology.  The KIDs themselves will likely be implemented in niobium nitride (NbN) microstrip, as described elsewhere in this conference proceedings \cite{glowacka2013development}.  Despite the W-band's astronomical importance and the use of other superconducting detectors such as transition edge sensors at these frequencies, there has been little work by the KID community in this area.  One of the reasons is the difficulties associated with coupling optical power at these wavelengths into a KID resonator.  The cut-off of frequency is aluminium (Al), which is commonly used as a sensing material for submillimetre-wave KIDs, has a gap frequency of around 90\,GHz.  This means it can only be used at the edge of the W-band, and here the performance is expected to be poor due to the proximity to cut-off.  In order to open up the whole of the band, an alternative sensor material or coupling arrangement is needed.  This issue is discussed in more detail in Section \ref{sec:optical_coupling}.  The target NEP for the devices is 4\,$\times$\,10$^{-18}$ W/$\sqrt{\text{Hz}}$, which would enable background-limited observing on the GLT.

\begin{figure}
\centering
\includegraphics[width=8cm]{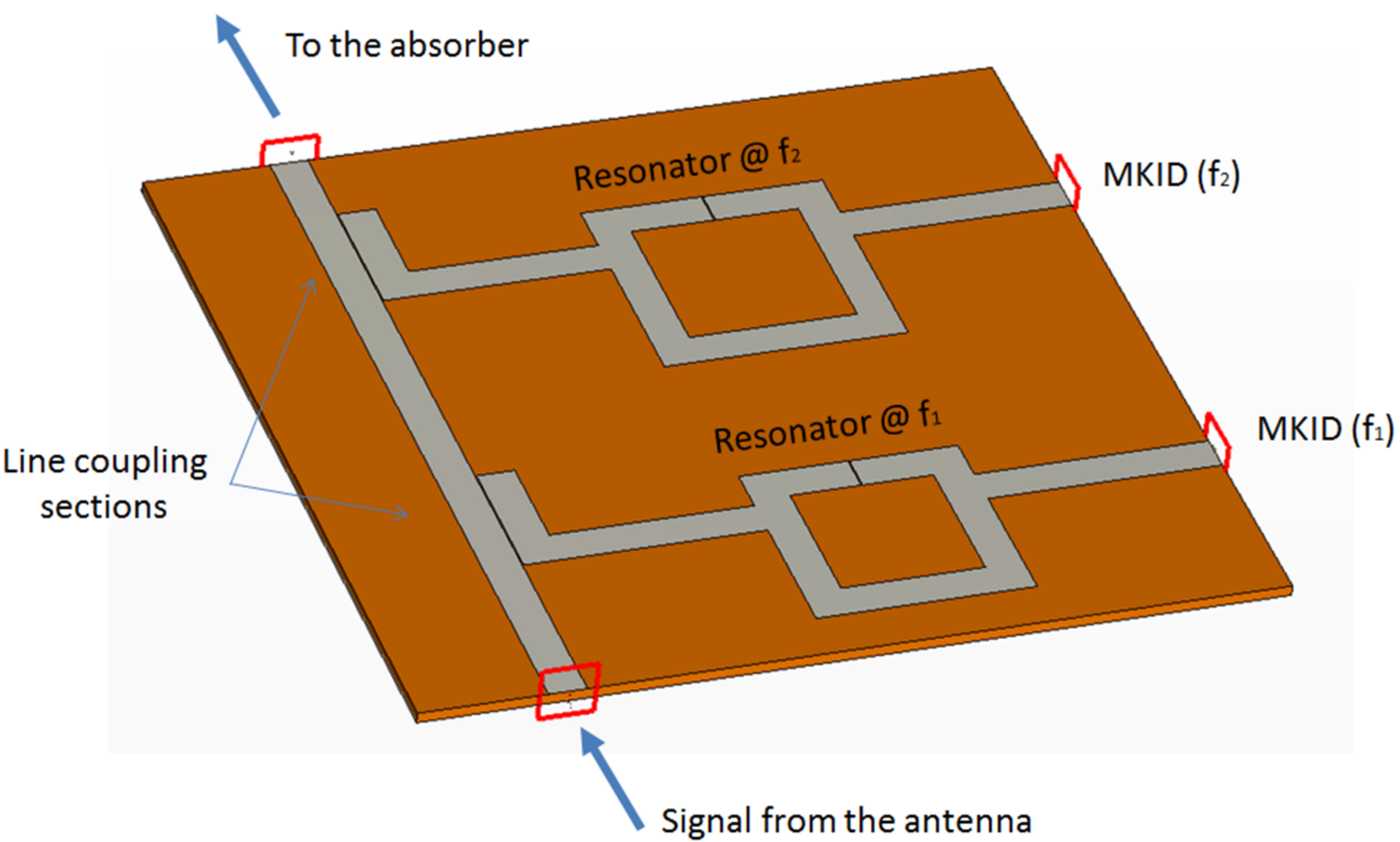}
\caption{Square open-loop RF filters for signal channelisation.  Additional poles are added to the response function by adding further loops.}
\label{fig:filter_designs} 
\end{figure}

CAMELS will use a horn antenna, and the waveguide probes will part of the detector chip.  The optical circuitry will also be implemented in NbN microstrip, which we expect to have extremely low Ohmic losses at W-band frequencies.  The target spectral resolution for the filter bank channels is R = 3000, which corresponds to a filter bandwidth of 3\,MHz.  We are investigating a range of filter implementation: one such design, the square open-loop resonator, is illustrated in Figure \ref{fig:filter_designs}.  In addition to the design of the individual filter elements, it also necessary to consider the frequency spacing of the filters.  It has been shown that by using a system of overlapping, highly interacting, filters, a larger fraction of the total power available in the bandwidth of the filter bank can be absorbed, compared with using non-overlapping channels \cite{kovacs2012superspec}.  However, it is not immediately clear how the interactions will effect the recovery of line parameters, or the stability of the system to manufacturing tolerances.  We are investigating an alternate design, whereby a polarisation splitting waveguide probe is used to feed a pair of filterbanks.  These filterbanks implement offset combs of filters, designed so that the two channels interlace to provide full coverage of the bandwidth.  The advantage of this design is it allows the filters in the individual polarisation channels to be spaced further apart in frequency, while still maintaining continuous coverage of the band.  This spacing is advantageous for two reasons.  The first is the interaction between adjacent filters elements is reduced, which simplifies design and makes the system more tolerant to manufacturing errors.  The second is that non-overlapping filters are insensitive to their physical spacing along the signal line, which may be useful if space on the chip proves a problem.  In order to compare different filter banks an appropriate performance metric is required, and our work in this area is discussed further in Section \ref{sec:filter_bank_performance}.  

CAMELS will use the canonical homodyne readout scheme for KIDs, as illustrated in \ref{fig:camels_system_diagram}.  A comb of tones, tuned to the readout point of each KID, will be generated at baseband (0--500\,MHz) using a programmable ADC card, upconverted to 1-5\,GHz and then used to probe the KIDs in the cryostat.  The signal transmitted through the KIDs is then amplified on the cold stage by a ultra-low noise HEMT, then once it has left the cryostat it is down-converted back to base-band and digitized by an ADC.  We plan to use a GPU-based system for the subsequent signal processing, rather than a dedicated digital-signal-processing board.  Similarly, we will implement the up and down-convertors using block microwave circuts, rather than custom printed circuit boards.  The rationale behind our design is to produce a simple readout system that we can build in-house, maintain at the telescope and can be modified easily to explore different readout schemes (this flexibility being granted by the GPU based system, which can be programmed with high-level languages).   

\section{Optical Coupling Schemes for W-Band KIDs}\label{sec:optical_coupling}

\begin{table}
\renewcommand{\arraystretch}{1.2}
\caption{Critical temperature $T_c$, pair-breaking frequency $\nu_c$ and normal state resistivity $\rho$ for common superconducting materials.  Data is for films deposited on silicon nitride, produced in the Cambridge Detector Physics group's processing facilities.}
\label{table:sc_gap_frequencies} 
\centering
\begin{tabular}{|c|c|c|c|}
	\hline
	Material & $T_c$ (K) & $\nu_c$ (GHz) & $\rho (20K)$ ($\times 10^{-8} \,\Omega m$) \\
	\hline
	\hline
	Nb & 9.2 & 671 & 8.8  \\
	Al & 1.2 &  88 &  0.56 \\
	NbN & 14.65 & 1068 & 245 \\
	Ta ($\beta$-phase) & 0.86 & 63 & 177  \\
	\hline
\end{tabular}
\end{table}

As noted in the previous section, an efficient scheme is needed to couple optical power into the KIDs over the whole of the W-band (70--90\,GHz).  In operation, a KID exploits the fact that a photon with sufficient energy can break a Cooper pair in a superconductor, producing quasiparticles.  These quasiparticles may be produced by direct absorption in the resonator, as in lumped element designs, or in a dedicated absorber from which they then diffuse into the resonator, as in antenna-coupled and high-energy devices.  The production of quasiparticles is then probed by measuring the corresponding changes in the resonant behaviour of the KID.  For incident photons to break Cooper pairs directly, we require
\begin{equation}\label{eqn:def_gap_frequency}
	h \nu > 2 \Delta,
\end{equation}
where $\Delta$ is the gap energy of the superconductor.  The operating temperature of a KID is normally well below the critical temperature $T_c$ of the superconductor, and so the gap energy is approximately the limiting value of the gap at absolute zero, $\Delta_0$.  For a BCS superconductor, the zero gap is related to the critical temperature $T_c$ of the superconductor by
\begin{equation}\label{eqn:bcs_relation}
	2 \Delta_0 = 3.5 k_b T_c, 
\end{equation}
and so combining (\ref{eqn:def_gap_frequency}) and (\ref{eqn:bcs_relation}), we find that the frequency $\nu_c$ at which the onset of pair-breaking is expected is given approximately by
\begin{equation}
	\nu_c = \frac{3.5 k_b}{h} T_c = 73 \, [\text{GHz/K}] \times T_c.
\end{equation}
In practice this represents a lower bound, and the signal frequency must exceed $\nu_c$ by a significant amount before absorption is observed.  Table \ref{table:sc_gap_frequencies} shows the pair-breaking frequencies of several common superconducting materials, based on $T_c$ measurements on films deposited at the Cavendish Laboratory.  

\begin{figure}
\centering
\includegraphics[width=7cm]{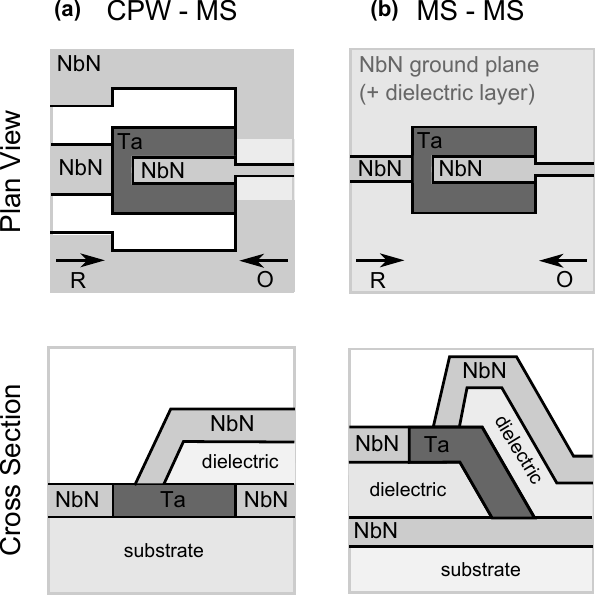}
\caption{\label{fig:proposed_optical_coupling_schemes}  Proposed optical coupling schemes.}
\end{figure}

Previously demonstrated antenna-coupled KIDs have typically used Al or mixed Nb-Al systems, with pair-breaking occuring in the Al \cite{day2006antenna, baryshev2011progress}.  However, as can be seen from Table \ref{table:sc_gap_frequencies}, this will not be possible for CAMELS, as $\nu_c$ lies in the middle of the W band.  Table \ref{table:sc_gap_frequencies} shows that instead $\beta$-phase Ta is an appropriate absorber.  $\beta$-phase Ta is, however, relatively lossy at GHz readout frequencies, and so we propose using it only for a small section at the shorted end of a quarter-wave microstrip resonator made from a different material, so as to not adversely degrade the resonator $Q$-factor. We will then run the feedline from the antenna over this lossy section to generate quasiparticles at the sensitive end of the resonator, in a similar manner to Day\cite{day2006antenna}.  This coupling scheme is illustrated for microstrip and CPW in Figure \ref{fig:proposed_optical_coupling_schemes}(a) and (b) respectively.

Care must be taken to ensure that there is no discontinuity in the impedance of the antenna microstrip at the point where it transitions onto the Ta absorber.  Any reflected power will decrease the optical efficiency.  Equivalently, an impedance mismatch in the resonator microstrip at the transition to Ta absorber will effect the resonant frequency and Q-factor.  The surface impedance of the line has a pronounced effect on the intrinsic impedance of the line, so a material is required with a normal state resistance that is similar to that of $\beta$-phase Ta (177 $\times$ 10$^{-8}$\,$\Omega$m).  For this reason we are intending to fabricate both the resonator and antenna microstrip from NbN, which has much higher normal state resistance (245 $\times$ 10$^{-8}$\,$\Omega$m) than the other resonator materials in Table \ref{table:sc_gap_frequencies}.  We have successfully fabricated NbN microstrip resonators, as detailed a paper by Glowacka in this conference proceedings\, \cite{glowacka2013development}.

\section{Assessing Filter Bank Performance}\label{sec:filter_bank_performance}

\subsection{The Fisher Information}

In addition to the detailed implementation of the individual filter elements, consideration must be given to the system level design of the filter bank, where there are many trade-offs.  Here we are using `system level' to refer to design choices such as the resolution, shape of each channel, filter-order and degree of overlap.  As an example, a question of interest is the degree to which the filter channels should be allowed to overlap/interact.  Highly interacting designs have been proposed that, summed over all channels, collect almost 100\% of the power incident across the band of operation \cite{kovacs2012superspec}.  Although the interaction increases the throughput, it comes at the expense of modification of filter shape, as well as the spreading of power optical power over several detector outputs.  The effect of these other factors on performance needs to be understood; for example, the changes in shape may complicate calibration, while the spreading of power will be disadvantageous if there is correlation between the noise in the detector outputs. 

What is needed to compare between different strategies is a performance metric that takes into account the intended use of the instrument, filter bank details and the noise in the system. The majority of planned on-chip spectrometers are intended to characterise galactic emission lines.  As such, their performance can be measured by the error with which --given a certain integration time-- they can recover line parameters such as the centre frequency, line width and total flux.  These errors also ultimately set the detection limits for lines.  

The approach we are taking for CAMELS is to use the Fisher information matrix \cite{riley2006mathematical} as a way of calculating expected values for these errors from the details of the design.  Consider a set of $N$ random variables $\{x_n\}$ that depend on a set of $M$ parameters $\{\theta_m\}$ according to the probability density function $P(\mathbf{x} | \boldsymbol{\theta})$.  The Fisher information matrix is defined by
\begin{equation}\label{eqn:def_fisher_matrix}
	\{ \mathsf{F} \}_{mn} (\boldsymbol{\theta})
	= \int \Bigl[ -\partial_{\theta_m} \partial_{\theta_n} 
	\ln P(\mathbf{x} | \boldsymbol{\theta}) \Bigr] 
	P(\mathbf{x} | \boldsymbol{\theta}) \, d^N \mathbf{x}.
\end{equation}
Its relevance to the current problem derives from the Cram\'er-Rao bound, which states that in an experiment where $\mathbf{x}$ is measured and used to recover $\boldsymbol{\theta}$, a lower bound on the covariance matrix $\mathsf{C}$ of the errors in the recovered parameters is given by
\begin{equation}\label{eqn:cramer_rao_bound}
	\mathsf{C} 
	= \bigl< \Delta \boldsymbol{\theta} \Delta \boldsymbol{\theta}^\dagger \bigr>
	= \mathsf{F}^{-1}.
\end{equation}
In the case of the filter bank spectrometer, with $\mathbf{x}$ representing the detector outputs and the $\boldsymbol{\theta}$ the line parameters, given a probabilistic model of the detector outputs we can use (\ref{eqn:cramer_rao_bound}) to calculate the instrument performance.  A qualitative argument for the Cram\'er-Rao bound is to note the contents of the square brackets is the derivative of the log-likelihood, the inverse of which is commonly used to estimate the covariance matrix of the model parameters in maximum-likelihood fitting (Laplace's method \cite{mackay2003information}).  From (\ref{eqn:def_fisher_matrix}) we see that $\mathsf{F}$ is simply the expected value of this derivative, so its inverse should also be the expected value of the covariance matrix. 

\subsection{Modelling a Spectrometer}\label{sec:spectrometer_model}

In the absence of noise, the we model the time-averaged spectrometer outputs as
\begin{equation}
	S_n (\boldsymbol{\theta}) = \int G_n(\nu) p(\nu | \boldsymbol{\theta}) \, d\nu,
\end{equation}
where $G_n(\nu)$ is the spectral response of the $n^\text{th}$ filter and $p(\nu | \boldsymbol{\theta})$ is the assumed line shape given a set of line parameters $\{\theta_n\}$.  We will assume the noise leads to a Gaussian error term described by the noise variance matrix $\mathsf{\Sigma}$, so that the likelihood of measuring a given set of outputs $\{D_n\}$ given the set of parameters is
\begin{equation}\label{eqn:spectrometer_likelihood}
\begin{aligned}
	& P(\mathbf{D} | \boldsymbol{\theta}) = \\ 
	& \frac{1}{(2\pi)^{N/2} || \mathsf{\Sigma} ||^{1/2}}
	\exp \Bigl( -\frac{1}{2} \bigl[\mathbf{D} - \mathbf{S}(\boldsymbol{\theta})\bigr]^\dagger
	\cdot \mathsf{\Sigma}^{-1} \cdot \bigl[\mathbf{D} - \mathbf{S}(\boldsymbol{\theta})\bigr] \Bigr).
\end{aligned}
\end{equation}
It is useful to briefly consider the likely possible forms that $\mathsf{\Sigma}$ will take:
\begin{itemize}
\item A simplified model of $\mathsf{\Sigma}$ is 
\begin{equation}
\begin{aligned}
	\Sigma_{mn} =& \, \bigl< \Delta D_m \Delta D_n \bigr> \\
	=& \, \Biggl[ \; \underbrace{\frac{\text{NEP}^2}{2 \tau}}_\text{(a)}
	+ \underbrace{ \frac{1}{\tau} 
		\int h \nu \, g_m(\nu) p_0(\nu) \, d\nu }_\text{(b)} \; \Biggr] 
		\, \delta_{mn} \\
	&+ \underbrace{ \frac{1}{\tau} 
	\int g_m(\nu) g_n(\nu) \bigl[p_0(\nu) \bigr]^2 \, d\nu }_\text{(c)},
\end{aligned}
\end{equation}
where $\tau$ is the integration time.  Term (a) represents the intrinsic detector noise, as characterised by the NEP.  Terms (b) and (c) represent photon noise contributions, with $p_0(\nu)$ the background loading.  Term (b) is the photon shot noise, and is uncorrelated between different detectors.  Term (c) represents classical radiometer noise (or `bunching' noise), and can lead to correlations in the noise at different outputs.
\item In the case where the internal detector noise dominates, $\mathsf{\Sigma}$ will be diagonal and will not depend on the filter profiles.
\item In the limit where the photon noise dominates there are two possibilities depending on whether the count or radiometric noise dominates, and in both cases $\mathsf{\Sigma}$ depends on the filter shapes.  When the photon noise dominates, $\mathsf{\Sigma}$ is diagonal.  When the radiometric noise term dominates, $\mathsf{\Sigma}$ will be non-diagonal if the filter channels overlap.
\item  The only time $\mathsf{\Sigma}$ will depend on $\boldsymbol{\theta}$ is if the photon noise due to the spectral line itself makes a significant contribution to the total noise.  This is exceedingly rare in astronomy, where the line signal is usually dwarfed by noise from the background loading.  For the rest of this note we will therefore assume $\mathsf{\Sigma}$ is independent of $\boldsymbol{\theta}$.
\end{itemize}
Given (\ref{eqn:spectrometer_likelihood}) The derivative of the log-likelihood is
\begin{equation}\label{eqn:deriv_log_likelihood}
\begin{aligned}
	\begin{aligned}
	& \partial_{\theta_i} \partial_{\theta_j} \log P(\mathbf{D} | \boldsymbol{\theta}) = \\ 
	& + \frac{1}{2} \partial_{\theta_i} \partial_{\theta_j} 
	\mathbf{S}(\boldsymbol{\theta})^T \cdot \Sigma^{-1} \cdot \mathbf{D}
	+ \frac{1}{2} \mathbf{D}^T \cdot \Sigma^{-1} \cdot 
	\partial_{\theta_i} \partial_{\theta_j} \mathbf{S}(\boldsymbol{\theta}) \\
	& -\frac{1}{2} \partial_{\theta_i} \partial_{\theta_j} \mathbf{S}(\boldsymbol{\theta})^T \cdot 
	\Sigma^{-1} \cdot \mathbf{S}(\boldsymbol{\theta})
	-\frac{1}{2} \mathbf{S}(\boldsymbol{\theta})^T \cdot 
	\Sigma^{-1} \cdot \partial_{\theta_i} \partial_{\theta_j} \mathbf{S}(\boldsymbol{\theta}) \\
	& -\frac{1}{2} \partial_{\theta_i} \mathbf{S}(\boldsymbol{\theta})^T \cdot 
	\Sigma^{-1} \cdot \partial_{\theta_j} \mathbf{S}(\boldsymbol{\theta})
	-\frac{1}{2} \partial_{\theta_j} \mathbf{S}(\boldsymbol{\theta})^T \cdot 
	\Sigma^{-1} \cdot \partial_{\theta_i} \mathbf{S}(\boldsymbol{\theta}).
\end{aligned}
\end{aligned}
\end{equation}
Noting that
\begin{equation}
	E \bigl[ \mathbf{D} \bigr] = \int \mathbf{D} \, P(\mathbf{D} | \boldsymbol{\theta}) \, d^n\mathbf{D}
	= \mathbf{S}(\boldsymbol{\theta}),	
\end{equation}
and that $\mathsf{\Sigma}$ is symmetric, it follows trivially that the Fisher information matrix is given by
\begin{equation}\label{eqn:spectrometer_fisher}
	\{\mathsf{F}\}_{mn} = 
	\bigr[ \partial_{\theta_m} \mathbf{S}(\boldsymbol{\theta}) \bigl]^\dagger 
	\cdot \mathsf{\Sigma}^{-1} \cdot 
	\bigr[ \partial_{\theta_n} \mathbf{S}(\boldsymbol{\theta}) \bigl].
\end{equation}
In conjunction with a line model and the Cram\'er-Rao bound, (\ref{eqn:spectrometer_fisher}) can be used to calculate the accuracy with which the spectrometer is expected to be able to recover a set of line parameters as a function of the position $\boldsymbol{\theta}$ of the line in the measurement space.  Its form is particular intuitive.  The terms in square brackets represent a responsivity - they measure how the outputs of the detectors are expected to change for changes in the line parameter.  The greater the `responsivity' for fixed noise, the smaller we would expect the errors in the recovered parameters to be.  Remembering that the error matrix is the inverse of $\mathsf{F}$, we see that this exactly what we would expect.  Similarly $\mathsf{F}$ is proportional to the inverse of the noise matrix $\Sigma$, so the error is expected to be proportional to the noise, in keeping with intuition.  At a more abstract level, (\ref{eqn:spectrometer_fisher}) can simply be regarded as rotating the noise covariance matrix into the basis of the model parameters in the limit of a small-signal model.

\subsection{Illustrative Example}\label{sec:example}

As a demonstration of the formalism introduced above we consider a reduced system.  We make the following assumptions:
\begin{itemize}
\item  We assume a galactic-like Gaussian line shape, parametrised by the integrated intensity $I_0$, the line-frequency $\nu_0$ and a line width $\Delta \nu$:
\begin{equation}\label{ex:line_shape}
	p(\nu | I_0, \nu_0, \sigma_L) = p_0(\nu) 
	+  \frac{2 I_0}{\Delta \nu \sqrt{2\pi}} 
	\exp \biggl[ -\frac{2(\nu - \nu_0)^2}{{\Delta \nu}^2} \biggr]
\end{equation}
\item We will initially assume $\Delta \nu$ is known a priori, although this is unlikely to be the case in a real measurement.
\item We assume the noise at the outputs is equal and uncorrelated, i.e.
\begin{equation}\label{eqn:ex_noise_matrix}
	\{ \mathsf{\Sigma} \}_{ij} = \sigma_N^2 \delta_{mn}
\end{equation}
\item We assume a simple set of top hat filters, as defined by
\begin{equation}\label{eqn:ex_filters}
	G_n(\nu) = 
	\begin{cases}
	1 & \bigl| \nu - \bigr[ n - \frac{1}{2} \bigl] \Delta f \bigr| \leq \frac{1}{2} \Delta f \\
	0 &  \text{otherwise}
	\end{cases}.
\end{equation}
\end{itemize}
The channel outputs for this model are given by
\begin{equation}
	S_m = \int\limits_{(m - 1) \Delta f}^{m \Delta f}
	\left[ p_0 (\nu) + \frac{2I_0}{\Delta \nu \sqrt{2 \pi}} 
	\exp \left[ -\frac{2 (\nu - \nu_0)^2}{{\Delta \nu}^2} \right] \right] \, d\nu,
\end{equation}
and the required derivatives with respect to the integrated flux and centre frequency are
\begin{equation}
\begin{aligned}
	\frac{dS_m}{dI_0}
	&= \frac{2}{\Delta \nu \sqrt{2 \pi}} \int\limits_{(m - 1) \Delta f}^{m \Delta f}
	\exp \left[ -\frac{2 (\nu - \nu_0)^2}{{\Delta \nu}^2} \right]\, d\nu \\
	&= N \Biggl( \frac{2 \bigl[ m \Delta f - \nu_0 \bigr]}{\Delta \nu} \Biggr)
	 - N \Biggl( \frac{2\bigl[ (m - 1) \Delta f - \nu_0 \bigr]}{\Delta \nu} \Biggr),
\end{aligned}
\end{equation}
and
\begin{equation}
\begin{aligned}
	\frac{dS_m}{d \nu_0} = \, 
		\frac{2I_0}{\Delta \nu \sqrt{2 \pi}} \Biggl[&
		\exp \Biggl( -\frac{2 \bigr[(m - 1) \Delta f - \nu_0 \bigl]^2}{{\Delta \nu}^2}\Biggr) \\
		&- \exp \Biggl( -\frac{2 \bigr[ m \Delta f - \nu_0 \bigl]^2}{{\Delta \nu}^2} \Biggr)
		\; \Biggr],
\end{aligned}
\end{equation}
where $N(x)$ is the cumulative distribution function of the standard normal distribution.  Defining the matrix $\mathsf{M}$ by 
\begin{equation}
\begin{gathered}
	\{\mathsf{M}(I_0, \nu_0) \}_{1m} = \frac{dS_m}{dI_0} \\
	\{\mathsf{M}(I_0, \nu_0) \}_{2m}= \frac{dS_m}{d \nu_0},
\end{gathered}
\end{equation}
and using (\ref{eqn:cramer_rao_bound}) and (\ref{eqn:spectrometer_fisher}), the Cram\'er-Rao bound on the covariance matrix is given by
\begin{equation}
\begin{aligned}
	\mathsf{C} (I_0, \nu_0) &= 
	\left( \begin{array}{cc} 
	\langle \sigma_{I_0}^2 \rangle & \langle \sigma_{I_0}^{} \sigma_{\nu_0}^{} \rangle \\
	\langle \sigma_{I_0}^{} \sigma_{\nu_0}^{} \rangle & \langle \sigma_{\nu_0}^2 \rangle 
	\end{array} \right) \\
	&= \sigma_N^2 \bigl[ \mathsf{M}^\dagger(I_0, \nu_0) \cdot \mathsf{M}(I_0, \nu_0) \bigr]^{-1}.
\end{aligned}
\end{equation}

\begin{figure}
\centering
\includegraphics[width = 7cm]{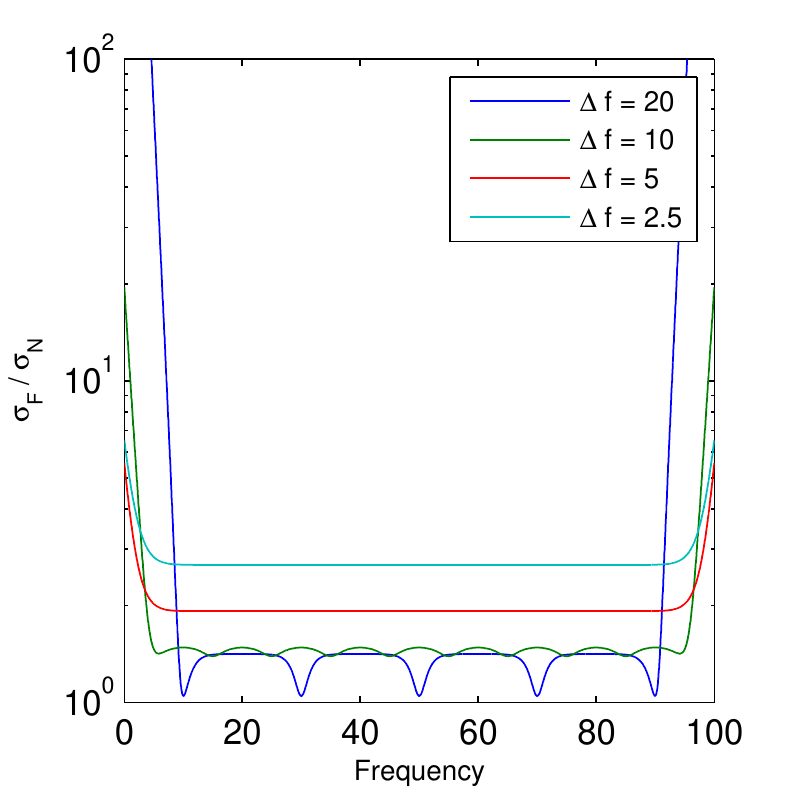}
\caption{\label{fig:filter_design_intensity} Plot showing the accuracy with which the total intensity of the line can be recovered as a function of frequency and filter width.  A line width of 10 frequency units is assumed.}
\end{figure}

\begin{figure}
\centering
\includegraphics[width = 7cm]{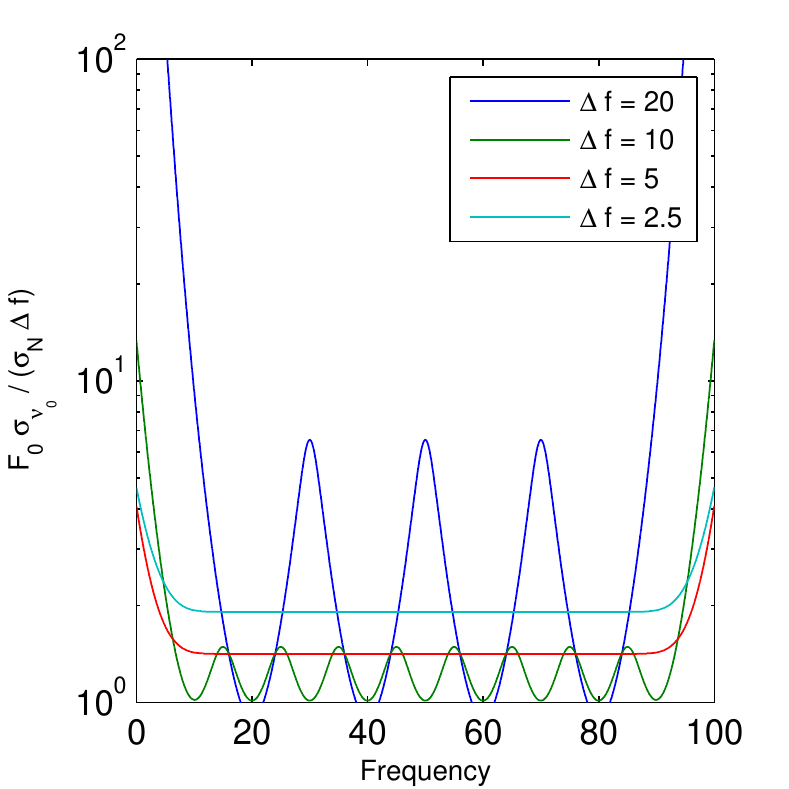}
\caption{\label{fig:filter_design_frequency} Plot showing the accuracy with which the centre frequency of the line can be recovered as a function of frequency and filter width.  A line width of 10 frequency units is assumed.}
\end{figure}

Figures \ref{fig:filter_design_intensity} and \ref{fig:filter_design_frequency} show the calculated Cram\'er-Rao bounds on the recovered integrated flux and centre frequency as a function of the centre-frequency, assuming a bank of non-overlapping top-hat filters that fills the bandwidth shown.  The different coloured lines show the results for different filter bandwidths, assuming a line width of 10 frequency units.  In each case the Cram\'er-Rao bound has been scaled so as to remove the dependence on the intensity of the line.  For $\Delta f$ $\leq$ 0.5\,$\Delta \nu$, the sensitivity of the instrument to the two parameters is uniform across the filter bank centre, but tapers off at the edge of the filter bank, as might be expected.  For $\Delta f$ $>$ 0.5\,$\Delta \nu$, there is significant ripple in the sensitivity across the band centre.  The sensitivity to the integrated flux peaks at the centre of each filter.  This can be explained intuitively by the fact at the filter edges the signal is shared between two noisy channels, giving a lower signal-to-noise ratio in each compared to if the signal were concentrated in one channel.  Conversely, the sensitivity to the position of the line peaks at the filter edge.  This is because in the centre of the channel, where the response is flat, the centre frequency can be changed by small amounts without affecting the output.  At the filter edges, changing the line position alters the balance of signal between the two channels, and this can be used to interpolate the line position to greater accuracy.  

This example is intended only as demonstration of the technique.  However, we see the power of the technique in its ability to put quantitative limits on performance.  For example, Figures \ref{fig:filter_design_intensity} and \ref{fig:filter_design_frequency} show clearly that we require $\Delta f \leq 0.5 \Delta \nu$ for uniformity of response across the centre of the bandwidth.  However, they also show that in practice we would probably not use a $\Delta f$ much smaller than 0.5 $\Delta \nu$, as the absolute sensitivity begins to suffer.  Although we may have been able to argue this behaviour intuitively in this reduced case, we would not have been able to a put a quantitative value of $\Delta f \approx 0.5 \Delta \nu$ on the optimal parameters without the technique.  In addition, the technique provides answers for much more complicated filter bank designs, with complicated filter shapes and overlap between channels, where the behaviour may not be as obvious.  There is no inherent increase in the complexity of the technique itself when applied to such designs.  The complexity arises in understanding the relationship between the many parameters of a more complex system and the overall performance, and it this, rather than any limitation of the technique itself, that is our primary reason for considering a reduced problem as an example.

\section{Conclusions}\label{sec:conclusions}

This paper has introduced the CAMELS instrument, which is intended as a technology pathfinder for on-chip spectrometer arrays at millimetre wavelengths.  The rationale behind the program is to place a prototype on a telescope, so as to investigate the operational issues associated with making well-calibrated science-grade observations with an IFBS.  It is planned that the instrument will observe in the frequency range 103.0--104.7\,GHz on the Greenland Telescope, starting in 2016.  Here we have outlined the system level design of the instrument, and discussed the issues associated with building a KID for W-band.  In addition, the use of the Fisher information as a performance metric for filter bank was discussed.

\bibliographystyle{IEEEtran}
\bibliography{isstt2013_bibliography}

\end{document}